

\input harvmac.tex

\def\e{\epsilon}
\def\ni{\noindent}
\def\bg{{\bf g}}
\def\bgt{{\bf g_{\tau}}}
\def\ttil{\tilde{\tau}}
\def\tri#1{\bigtriangleup( #1 )}
\def\a{\alpha}
\def\b{\beta}
\def\g{\gamma}
\def\l{\lambda}
\def\L{\Lambda}

\def\lt#1{\l_{\tau^{#1}(0)}}
\def\qp#1{q\left( {{#1}h \over k} \right)}
\def\s{\sigma}
\def\EM{\hat{E}_M}
\def\fiz{\hat{F}^i(z)}
\def\rltj{|\Lambda_{\tau(j)}>}
\def\rlj{|\Lambda_j>}

\def\lltj{<\Lambda_{\tau(j)}|}
\def\llj{<\Lambda_j|}


 \nref\OTc{D.I. Olive and N. Turok, {\it Nucl.Phys.} {\bf B257
[FS14]} (1985) 277-301, \lq\lq Local Conserved Densities and
Zero Curvature Conditions for Toda Lattice Field Theories"}

\nref\W{G. Wilson, {\it Ergod. Th. \& Dynam. Sys.} {\bf 1}
(1981) 361-380, \lq\lq The modified Lax and two-dimensional
Toda lattice equations associated with simple Lie algebras"}

\nref\OTUa{ D.I. Olive, N. Turok and J.W.R. Underwood, {\it
Nuclear Physics} {\bf B401} (1993) 663-697,``Solitons and
the Energy-Momentum Tensor for Affine Toda Theory"  }

\nref\OTUb{D.I. Olive, N. Turok and J.W.R. Underwood, {\it
Nuclear Physics} {\bf B409} (1993)  509-546,  ``Affine Toda
Solitons and Vertex Operators"  hep-th/9305160 }

\nref\KOa{ M.A.C. Kneipp and D. I. Olive , {\it Nucl. Phys.}
{\bf B408} (1993) 565-578, ``Crossing and Antisolitons in
Affine Toda Theories"  hep-th/9305154  }

\nref\FLO{A. Fring, H.C. Liao and D.I. Olive, {\it Phys.
Lett.} {\bf B 266} (1991) 82-86, \lq\lq The Mass Spectrum and
Coupling in Affine Toda Theories"}

\nref\FO{ A. Fring and D.I. Olive,  {\it Nucl. Phys.} {\bf
B379} (1992) 429-447, \lq\lq The Fusing Rule and the
Scattering Matrix of Affine Toda Theory"}

\nref\Fre{M.D.Freeman, {\it Phys. Lett.} {\bf B 217} (1991)
57, \lq\lq On the Mass Spectrum of Affine Toda Field Theory"}

\nref\Dora{P.E. Dorey, {\it Nucl. Phys.} {\bf B358} (1991)
654,\lq\lq Root Systems and Purely Elastic S-Matrices"}

\nref\Dorb{P.E. Dorey, {\it Nucl. Phys.} {\bf B374} (1992) 741
\lq\lq Root Systems and Purely Elastic S-Matrices 2"
hep-th/9110058 }

\nref\Kac{ V.G. Kac, \lq\lq Infinite Dimensional Lie
Algebras", Cambridge University Press, Third Edition, 1990.}

\nref\OTa{D.I Olive and N. Turok, {\it Nucl. Phys.}  {\bf B215
[FS7]} (1983) 470-494, \lq\lq The Symmetries of Dynkin
Diagrams and the Reduction of Toda Field Equations"}

\nref\BCDSb{ H.W. Braden, E. Corrigan, P.E. Dorey and R.
Sasaki, {\it Nucl. Phys.} {\bf B338} (1990) 689: \lq\lq Affine
Toda Field Theory and Exact S-Matrices".}

\nref\Spr{T. A. Springer, {\it Inv. Math.}{\bf 25} (1974) 159
, ``Regular elements of finite reflection groups". }

\nref\Dorc{P.E. Dorey, \lq\lq Hidden geometrical structures
in integrable models" in ``Integrable Quantum Field Theories"
ed. L. Bonora, G. Mussardo, A Schwimmer, L. Girardello and M.
Martellini, Plenum 1993, 83-97 hep-th/9212143 }

\nref\Dord{P.E.Dorey, {\it Phys. Lett.} {\bf B312}
(1993)291-298, ``A Remark on the coupling dependence in Affine
Toda Field Theories"  hep-th/9304149 }

\nref\FKa{A. Fring and R. Koberle,`On exact S-matrices for
non-simply laced Affine Toda Field Theories'
USP-IFQSC/TH/93-13}

\nref\ZamA{A.B. Zamolodchikov, {\it Int. J. Mod. Phys.} {\bf
A3} (1988) 743-750, \lq\lq Integrals of Motion in scaling
3-State Potts Model Quantum Theory"}

\nref\ZamB{A.B. Zamolodchikov, {\it Int. J. Mod. Phys.} {\bf
A4} (1989) 4235-4248, \lq\lq Integrals of Motion and S-Matrix
of the (Scaled) $T=T_c$ Ising Model with Magnetic Field"}

\nref\ZamC{A.B. Zamolodchikov , \lq\lq Integrable Field Theory
from Conformal Field Theory", in \lq\lq Integrable Systems in
Quantum Field Theory and Statistical Mechanics", {\it
Advanced Studies in Pure Mathematics"} {\bf 19} 1989 641-674
(Academic Press) }

\nref\Hola{T.J. Hollowood, {\it Nucl. Phys.} {\bf B384} (1992)
523-540, \lq\lq Solitons in Affine Toda Field Theories"}

\nref\DGZ{G.W. Delius, M.T. Grisaru and D. Zanon, {\it Nucl.
Phys.} {\bf B382} (1992) 365, \lq\lq Exact S-matrices for
non-simply laced affine Toda theories" hep-th/9112007}

\nref\CDS{E. Corrigan, P.E. Dorey and  R. Sasaki: \lq\lq On a
Generalised Bootstrap Principle",  Preprint DTP-93/19,
YITP/U-93-09  hep-th/9304065}

\nref\MM{N.J.MacKay and W.A.McGhee, {\it Int. J. Mod. Phys.}
{\bf A8} (1993) 2791-2808;
 \lq\lq Affine Toda Solitons and Automorphisms of Dynkin
Diagrams" hep-th/9208057 }

\nref\ACFGZ{H.Aratyn, C.P. Constantinidis, L.A. Ferreira, J.F.
Gomes and A.H. Zimerman, {\it Nucl. Phys.} {\bf B406}
(1993)727-770,
\lq\lq Hirota's Solitons in the Affine and the Conformal
Affine Toda Models",  hep-th/9212086 }


\Title{\vbox{\baselineskip12pt\hbox{Swansea SWAT/93-94/19}
\hbox{hep-th/9404030} }} {\vbox{\centerline{Solitons and
Vertex Operators in}\centerline{Twisted Affine Toda Field
Theories}}}


\centerline{Marco A.C. Kneipp and David I. Olive}

\centerline{\it{Department of Mathematics, University College
of Swansea,}}
\centerline{\it{Swansea SA2 8PP, Wales, UK.}}

\vskip 13mm
\centerline{\bf Abstract}

Affine Toda field theories in two dimensions constitute
families of integrable, relativistically
invariant field theories in correspondence with the affine
Kac-Moody algebras. The particles which are the quantum
excitations of the fields display interesting patterns in
their masses and coupling and which have recently been shown
to extend to the classical soliton solutions arising when the
couplings are imaginary. Here these results are extended from
the untwisted to the twisted algebras. The new soliton
solutions and their masses are found by a folding procedure
which can be applied to the affine Kac-Moody algebras
themselves to provide new insights into their structures. The
relevant foldings are related to inner automorphisms of the
associated finite dimensional Lie group which are calculated
explicitly and related to what is known as the twisted
Coxeter element. The fact that the twisted affine Kac-Moody
algebras possess vertex operator constructions emerges
naturally and is relevant to the soliton solutions.

\Date{March 1994} 


\newsec{Introduction}

Affine Toda theories in two dimensions
are integrable and possess an infinite number of local
conservation laws \OTc \W whose charges generate what can be
considered as an infinite dimensional Poincar\'e algebra,
\eqna\eIi
$$\eqalignno{\left[ P^{(M)} , P^{(N)} \right] & =  0 , &\eIi
a\cr
             \left[ K , P^{(M)} \right] & = i M P^{(M)} . &
\eIi b\cr }$$
The Lorentz boost, $K$, measures the Lorentz spin, $M$, of
the  ``momentum", $P^{(M)}$. The values of the integers $M$
for which the momenta  $P^{(M)}$ are non zero form the set
of  exponents of the associated affine Kac Moody algebra whose
root system appears in the original equations of motion. The
affine Toda field theory possesses critical points with
$W$-symmetry  and the symmetries \eIi{}\  can be regarded
as the relics of this which survive when the critical theory
is deformed  in the appropriate integrable manner.

A remarkable but well known mathematical fact is that the
underlying affine Kac Moody algebra possesses a subalgebra
called the ``principal Heisenberg subalgebra"  which, when
augmented by the ``principal derivation",  is precisely
isomorphic to  \eIi\ at level zero. Recent
studies \OTUa \OTUb \KOa\ of the classical soliton solutions
of the affine Toda field theories for imaginary couplings
(thereby extending the familiar  theory of sine-Gordon
solitons) have revealed a formulation in which the above
mentioned subalgebra does indeed act on the solutions as the
Poincar\'e algebra \eIi{} . It is this coupling of space-time
and  internal symmetries which essentially
explains  why so many of the space time properties of the
particles and solitons of the theory (masses, coupling and
scattering matrices) possess a Lie algebraic nature
\OTUa  --\Dorb  . However this formalism has so far included
only the untwisted affine Kac Moody algebras and not the
twisted ones.

The untwisted algebras has been traditionally easier to
understand and, up to now, found more physical applications.
The twisted algebras can, in any case, be understood as
subalgebras of the simple simply-laced untwisted ones. In
fact there are two natural but distinct ways of achieving
this. One way has been well studied in the mathematics
literature\Kac\  whereas the other has been found to be
useful in
the physics literature dealing with affine Toda field
theories\OTa \BCDSb .

Here we shall find it worthwhile relate these two hitherto
distinct procedures.
{}From the mathematical point of view we shall illuminate the
``twisted Coxeter element"\Spr\ which plays a r\^ole in the
grading of the twisted algebra, as well as the vertex
operator construction which is valid for the twisted algebras
despite the fact that the Dynkin diagrams conventionally
associated with them are not simply laced.

More physically, the results will clarify both the spectrum
and the coupling of the particles of the twisted affine Toda
field theories. Moreover the general formalism for the
soliton solutions will be found to extend in a
straightforward manner.
Recent observations of Dorey\Dorc \Dord\  will be understood
better and set in a
more  general framework. Part of the interest of the twisted
theories is that they constitute a  different  sort of
integrable deformation of the conformal Toda theories as
compared  to the untwisted theories, in the sense of
Zamolodchikov\ZamA \ZamB \ZamC\ .

Section 2 reviews the usual classification of the twisted
affine Kac-Moody algebras and the Dynkin diagrams associated
with them as explained by  Kac in his book. In the second
construction these are obtained by ``folding" an untwisted
simply laced extended Dynkin diagram
$\bigtriangleup
\left(\bg^{(1)} \right)$  making use of a special symmetry of
$\bigtriangleup
\left(\bg^{(1)}\right)$  which has the property that it can
be lifted to an inner automorphism of the finite dimensional
Lie algebra
$\bg$. Such  symmetry forms the finite group \OTa \OTUb
\eqn\eW{W_0(\bg)\cong  W(\bg) \cap
Aut\bigtriangleup\left(\bg^{(1)}\right)
\cong Z(G)}
where $W(\bg)$ is the Weyl group of $\bg$, and $Z(G)$ the
centre of the simply connected Lie Group, $G$, whose Lie
algebra is $\bg$. All twisted Kac-Moody algebras, except
${\bf A}_{2n}^{(2)}$, can be so obtained in an essentially
unique way. ${\bf A}_{2n}^{(2)}$ is exceptional in having a
Dynkin diagram with  three different root lengths and  needs
special treatment in both approaches and so will be ignored
as we seek general arguments.

These folding procedures are also applied to the
corresponding affine Toda field theory  equations and results
concerning masses and solutions are immediately deduced.

Section 3 presents the main ideas in rather general form. The
lift  of any element $\tau \in W_0$, \eW,  to  an inner
automorphism of $\bg$ is considered. Its fixed point set,
$\bgt$, is defined, and shown to be acted upon a natural way
under conjugation by $S$, the lift of the Coxeter element of
$\bg$. Thus $S$ acts as an outer automorphism of
$\bgt$ which will be related to the twisted Coxeter element of
the semisimple part of
$\bgt$ when this is simple. It is further shown that if
$\tau$ has  order $k$, then
$S^k$ lies in the Lie group $G_{\tau}$ itself and this
observation provides
a crucial link with the alternative construction described by
Kac.

Section 4 describes a concrete expression for the group
element, $S$,  conjugation by which yields the inner
automorphism
of $G$ corresponding to $\tau$ in the cases
$\bg$ is simply laced and $\tau $ is direct. From this is
deduced the precise structure of
$G_{\tau}$. As this  has to have the same rank as $G$  it
is unclear how it  is related  to the fold of
$\bigtriangleup \left(\bg^{(1)}\right)$. It turns out that
$G_{\tau}$ is never semisimple as it possesses $(k-1)$
invariant U(1) factors. When the remaining semisimple factor
is actually simple, we recognize a relationship to the
twisted Dynkin diagrams appearing in the classification
decribed by Kac \Kac . The remaining
possibilities are interesting but were not considered in \Kac .

The construction of section 4 has several geometrical
features which are described in section 5 and used to view
the root system of the simple part of $\bgt$ as being acted
upon by  a twisted Coxeter element following from the action
of
$S$ mentioned above.

We then see how a basis for the twisted affine Kac Moody
algebra can be  formulated in terms of a twisted principal
Heisenberg subalgebra and the  associated quantities
ad-diagonalising them. Viewed this way, the basis for the
twisted algebra is simply a subset of the corresponding basis
for
$\bg^{(1)}$, the untwisted simply-laced affine Kac-Moody
algebra. As a consequence it is shown to
inherit its vertex operator construction.

In section 6 we discuss the application of these results to
understanding the properties  of the particles and solitons
in the twisted affine Toda theories. In particular we find
that there is a sense in which the energy momentum tensor is
unchanged by folding. As a consequence, the twisted  soliton
mass spectrum is a subset of the spectrum of the unfolded
$\bg^{(1)}$
theory, in the same manner as for the spectrum of masses of
the quantum field excitations.

For completeness, we also discuss how our general results
apply to the other kind of folding,
namely that which yields the untwisted non-simply laced affine
Kac-Moody algebras. Here the results
for the two kinds of mass spectrum differ and our method
gives a simple explanation of this,
thereby confirming previous results.

\newsec{Folding and twisted theories}

The conventional notation for the  twisted affine Kac-Moody
algebra is the designation ${\bf X}_n^{(k)}$, $k\geq 2$.  As
explained in Kac's book\Kac ,
${\bf X}_n$ denotes a simple, simply laced Lie algebra of
rank
$n$ endowed with a diagram automorphism (of the Dynkin
diagram
$\bigtriangleup \left({\bf X}_n \right)$ ) of order $k$.  It
turns out that the construction can be summarised by the
statement that the Dynkin diagram $\bigtriangleup \left(
{\bf X}_n^{(k)} \right)$ has  as Cartan matrix
$K(({\bf Y^v})^{(1)})^T$ where ${\bf Y}$ is the subalgebra of
${\bf X}_n$ fixed by the automorphism of ${\bf X}_n$ which is
the lift of the diagram automorphism; ${\bf Y^v}$ is its
dual, that is the algebra with the roots and coroots
interchanged. Thus
$K(({\bf Y^v})^{(1)})^T$ is the transpose of the extended
Cartan  matrix of ${\bf Y^v}$. Following Kac's book, table 1
lists the possibilities with exception of ${\bf
A}_{2n}^{(2)}$ which has be treated as a special case in
both the conventional approach and in ours to follow.

In table 1 the vertices of $\bigtriangleup \left(
{\bf X}_n^{(k)}
\right)$  are numbered for future reference. Note that the
integer $l$ is chosen so that the folded Dynkin diagrams
$\bigtriangleup ({\bf D}_{l+1}^{(2)} )$  and
$\bigtriangleup ({\bf A}_{2l-1}^{(2)} )$ each possess
precisely $(l+1)$ vertices.

In considering affine Toda field theories, it has been found
helpful to view the diagrams in table 1 as arising in a
different way, namely by ``folding"\OTa\  the extended Dynkin
diagrams of simply laced simple Lie algebras. Thus, by virtue
of a symmetry of the extended Dynkin diagram $\bigtriangleup
({\bf D}_{2l}^{(1)} )$ it can be folded to
$\bigtriangleup ({\bf A}_{2l-1}^{(2)} )$ by identifying
points related by the symmetry. It is sufficient to denote
these symmetries as a permutation of the tip points of the
extended Dynkin diagram, namely those symmetrically related
to the vertex
$(0)$. Each diagram in table 1 can be found as indicated in
table 2. The symmetry to be used for the folding is
specified in the last column, using the numbering of
vertices of $\bigtriangleup (\bg^{(1)})$ and the conventional
notation for permutations.

Notice that the automorphism of $\bigtriangleup (
\bg^{(1)})$ needed to obtain the diagrams of table 2 possess
two important properties: they are direct (i.e. never relate
two linked vertices) and they are elements of $W_0(\bg)$, \eW
. This means that unlike the element of Aut$\bigtriangleup
({\bf X_n })$ used in the construction of table 1, as
described by Kac, these
automorphisms of
$\bigtriangleup \left(\bg^{(1)} \right)$ can be lifted to
inner automorphisms of the Lie algebra $\bg$ (as well as
the corresponding Lie
group
$G$). This inner automorphism will be important in what
follows and will be constructed explicitly. It has been
shown that\OTa \OTUb
\eqn\eWW{W_0(\bg) \cong Z(G),}
where $Z(G)$ is the centre of the simply connected Lie
group G whose Lie algebra is $\bg$. The result \eWW\ makes
it easy to scan all the possibilities for elements of
$W_0(\bg) $ when $\bg$ is simple and simply laced. We find
that two possibilities remain beyond those in table 2.
$W_0({\bf D}_{2n+1})$ possesses an element of order 4 not
listed
above. It is, however, not ``direct" and so excluded from
most of
our argument. The remaining case turns out to be of
considerable interest,
\eqn\eWA{W_0({\bf SU(N)}) \cong W_0({\bf A}_{N-1}) \cong
Z_N.}
If $N=mn$ , we have $\tau \in W_0$ defined by
\eqn\et{\tau(i) = i +
n (\hbox{ mod }mn),}
using the numbering of vertices in table 3. As $\tau^m(i) =
i+mn = i$, this has order
$m$. In this case, ``folding" gives
$\bigtriangleup({\bf A_{n-1}}^{(1)})$.

We shall now explain why the folding procedure of table 2 is
useful in affine Toda field theory and deduce the form of
the general soliton solution for the twisted theories,
generalising those of the untwisted theories.

Following the treatment of \OTUb , let $<i>$ denote the set of
vertices of
$\bigtriangleup(\bg^{(1)})$ related to the vertex $i$ by
the action of $\tau \in $\ Aut$\tri{\bg^{(1)}}$, that is, its
orbit. Then, if $\tau$ is direct, so that no two points of
$<i>$ are ever linked directly,
\eqn\eK{K_{<i><j>} = \sum_{j \in <j>} K_{ij} }
defines a new, folded, Cartan matrix. This is the precise way
in which the
folded diagrams of Table 1 were found by folding the diagrams
in Table 2, using the specific
element $\tau$ listed in the last column. The set of coprime
integers $m_i$, defined as
$\sum_{j=0} K_{ij}m_j = 0$, obviously fold according to
\eqn\em{m_{<i>} = m_i }
as $\sum_{<j>} K_{<i><j>}m_{<j>} = 0$, and the $m_{<i>}$
remain
coprime integers.

If we define the variables
\eqn\ephi{\phi_i = \a_i . \phi , \hskip 2cm i=0,1,\ldots,r}
where $(\a_0, \a_1, \ldots ,\a_r )$ are an extended set of
simple roots, and
\eqn\eKK{K_{ij}={2\alpha_i\cdot\alpha_j \over \alpha_j^2},}
we have the relations
\eqn\eefi{\sum_{i=0}^r {2 m_i \over \a_i^2} \phi_i = 0,
         \hskip 2cm
           \sum_{i=0}^r {2 m_i \over \a_i^2}  K_{ij} = 0.}

The advantage of introducing a redundant variable is that
the affine Toda field equation of motion
associated with the Cartan matrix $K_{ij}$ can be written
\eqn\eem{\partial^2 \phi_i + {\mu^2\over\beta}\sum_{j=0}^r
         K_{ij}m_je^{\beta\phi_j} = 0 ,
\hskip 1cm i = 0, 1, \ldots , r , }
subject to \eefi , and that this version better displays the
symmetry of the Dynkin diagram whose
Cartan matrix is
$K$.  This can be done whether the algebra is twisted or not.
Equation \eem\ can
be linearised to
\eqn\eqm{\partial^2 \phi_i + \mu^2 \sum_{j=0}^r  K_{ij}m_j
        \phi_j = 0. }
Thus the squared masses of the quantum particles excitations
of the fields equal $\mu^2$ times the eigenvalues of the
matrix $C_{ij}' = K_{ij}m_j$ which is similar to $C_{ij} =
m_iK_{ij}$, so sharing the same eigenvalues (cf \OTUb ) .
Because
of the condition \eefi , the eigenvalue 0 is excluded leaving
precisely $r$ values.

The basic result concerning the folding of equation \eem\ via
\eK\ is that if $\phi_{<i>}$ is a solution of the folded
equation, with $C'_{<i><j>}$ replaced by $C'_{ij}$ in \eem
, then
\eqn\efi{\phi_i = \phi_{<i>} }
is a solution to the unfolded equation that displays the
symmetry $\phi_i = \phi_{\tau_{(i)}}$. Conversely, all such
symmetric solutions furnish solutions via \efi\ to the folded
equations.

In particular, the mass spectrum of the quantum field
excitation  particles of the
folded theory automatically forms a subset of the unfolded
mass
spectrum. These results holds for any diagram automorphism
$\tau
\in$ Aut$\tri{\bg^{(1)}}$ as long as it is direct. The group
Aut$\tri{\bg^{(1)}}$ is actually a semidirect product of two
subgroups Aut$\tri{\bg}$ and $W_0(\bg)$ (\OTa \OTUb ). Both
these subgroups are relevant. Folding with $\tau \in $
Aut$\tri{\bg}$ yields a non simply laced untwisted affine Kac
Moody algebra. Such $\tau$ can be lifted to an outer
automorphism of the Lie algebra $\bg$, preserving the
principal ${\bf SO(3)}$ subalgebra. All   this was discussed
as in
detail in \OTUb  and will not be pursued again here.
Instead, as explained above, we consider the effect of
elements  of
$W_0(\bg)$. These can be lifted to inner automorphisms of
$\bg$ which do not preserve the principal
${\bf SO(3)}$ subalgebra and fold to give twisted algebras.

In untwisted affine Toda theories,  the quantum field particle
excitations
are in one to one correspondence with points of the ordinary
Dynkin diagram $\tri{\bg}$. Mass degeneracies reflect the
symmetries of the diagram and are broken by foldings by
$\tau \in$ Aut$\tri{\bg}$. We have just seen that for $\tau
\in W_0$ a subset of particles survive folding and one of
our results will be to determine precisely which in terms of
a simple general formula. The surviving particles
correspond to the subset of vertices first found  in the
explicit
calculations of Braden et al\BCDSb , and rederived in section
6 to
follow.

The general soliton solution to \eem\ for imaginary coupling
$\beta$ can be written
\eqn\esoli{e^{-\beta\phi_i} = \prod_{j=0}^n (M_j)^{K_{ij}} ,}
where
\eqn\ef{M_j = \langle \Lambda_j | g(t) |\Lambda_j \rangle }
is a tau function, obeying an expectation value with respect
to the $j^{th}$ fundamental highest weight state of the
algebra. $g(t)$  is a group element constructed in a
specific way from the soliton data. For the untwisted
theories \esoli\ is derived by a simple rearrangement of the
general expression of \OTUa . We shall now show that, unlike
the formula of \OTUa , \esoli\ has the virtue of applying to
the twisted affine Toda theories as well.

First note that, if  $M_j$ symmetric in the sense $M_j =
M_{\tau(j)}$, we can define  $M_{<j>}$  to be equal to it and
use
\eK\ and \efi , to indeed find
\eqn\esolit{e^{-\beta\phi_{<i>}} =
\prod_{<j>=<0>}(M_{<j>})^{K_{<i><j>}} .}
It will be verified in section 6 as a result of our
construction that indeed $M_j = M_{\tau(j)}$ in the twisted
Kac-Moody algebra.

\newsec{Action of $W_0(\bg)$ on the principal Coxeter
element $S$}

The first, and main, part of our argument is quite simple
and general. It applies to any element of $\tau \in
W_0(\bg)$,
\eW , whether or not $\bg$  is simply laced, and whether or
not $\tau$ is direct.

Recall that any element $\tau$ of $W_0(\bg)$ is uniquely
characterised by its action, $\tau(0)$, on the vertex 0 of
the extended Dynkin diagram, $\tri{\bg^{(1)}}$, whose
deletion  yields the ordinary Dynkin diagram $\tri{\bg}$
\OTUb . Moreover if
$\tau \neq 1$ it moves each tip point of $\tri{\bg^{(1)}}$.
Suppose $\tau \in W_0(\bg)$ has order $k$:
\eqn\etk{\tau^k = 1.}
Because $W_0(\bg) \subset W(\bg)$, the Weyl group of $\bg$,
$\tau$ can be lifted to an inner automorphism $\ttil $ of
$\bg$:
\eqn\etau{\ttil (p) = TpT^{-1}, \hskip.5cm p\in \bg, \; \;
T\in H' \subset G  ,\hskip.5cm }
where, as stated $T$ lies in $H'$ the maximal torus of
$G$ whose Lie algebra is the Cartan subalgebra in
apposition, ${\bf h}'$. This is because
\eqn\ete{\ttil(E_1) = E_1, \hskip 2cm E_1 =
\sum_{i=0}^{r} \sqrt{m_i} E_{\alpha_i} }
and $H'$ is the centraliser of $E_1$. From \etk\
follows that
\eqn\extra{T^k \in Z(G).}
where $Z(G)$ is the centre of the simply connected Lie
group G whose Lie algebra is $\bg$.

The
definition \etau\ of $\ttil$ leaves some ambiguity in $T$,
to be discussed later, but, irrespective of this,  it was
shown that\OTUb
\eqn\etsts{TST^{-1}S^{-1} = e^{-2\pi i \l_{\tau(0)}^v . H}
\equiv z(\tau),}
where $S$ is the principal  element of $G$
\eqn\es{S = \exp {2\pi i T^3\over h(\bg)},}
and $\lambda_i^v=2\lambda_i/\alpha_i^2$ is a fundamental
coweight.  The angular momentum $T^3$ lies in
the intersection of the principal ${\bf SO(3)}$ subalgebra
of
$\bg$ with the original Cartan subalgebra ${\bf h}$ of
$\bg$.
$h(\bg)$ denotes the Coxeter number of $\bg$. As
\eqn\ess{S^{h(\bg)} \in Z(G),}
conjugation by $S$ grades $\bg$ into $h(\bg)$ eigenspaces
\eqn\egra{\bg = \bg_0 \oplus \bg_1 \oplus \cdots \oplus
\bg_{h({\bf g}) -1}
\hskip 1cm{\rm where}\hskip 1cm S \bg_m S^{-1} =e^{2\pi i
m \over h(\bg)} \bg_m ,}
and therefore
\eqn\egraa{\bg_0 = {\bf h} ; \hskip 2cm E_1 \in \bg_1.}
The known form of the zero curvature potentials of the affine
Toda field theories indicates that
these  eigenspaces are of crucial importance in understanding
the integrability of these
theories.

The group element $z(\tau)$ in \etsts\ lies in $Z(G)$.
Thus \etsts\ is a key equation, stating that $T$ and $S$
almost commute, despite the fact that they lie in maximum
tori in apposition. It follows from \etsts\ that, for any
pair of integers $m$ and $n$,
\eqn\ets{T^m S^n = z(\tau )^{mn} S^n T^m .}
Our first deduction is that $z(\tau)$ has precisely the same
order, $k$, \etk\ , as $\tau$ and that $k$ necessarily
divides the Coxeter number $h(\bg)$.

By \extra\ and \ets , $z(\tau)^k = 1$, but if $z(\tau)$ had
a lower order,
$k'$, then $T^{k'}$ would commute with $S$, and so, by
\etau\ and \egra , would lie in the intersection of the two
maximal tori in apposition with each other, $H $ and
$ H'$. As this intersection is simply $Z(G)$, this
would imply $\tau$ had order $k'$, contrary to the
hypothesis. It now follows that $k$ provides the smallest
power of $S$ that commutes with $T$. By  \ess\ it then
follows that $k$ divides $h(\bg)$.

Now let us define $G_{\tau}$, the fixed point subgroup of
$G$ with respect to $\ttil$,
\eqn\eIIIix{G_{\tau} = \{a \in G ; \ T a T^{-1} = a\} . }
Since $z(\tau)^k = 1$, from \ets\ we see that $S^k$
commutes with $T$ and hence lies in
$G_{\tau}$ while
$S$ itself does not when $k\geq 2$. Nevertheless
\eqn\eIIIx{S a S^{-1} \in G_{\tau}  \ {\rm for\ all}\ a
\in G_{\tau}}
as $T$ commutes with $SaS^{-1}$ by virtue of \etsts\ and
\eIIIx . Thus conjugation by $S$ produces an outer
automorphism of
$G_{\tau}$, with the property that its $k^{th}$ power is
inner.

The remaining argument is to relate this action of $S$ to
that of the so called twisted Coxeter element. The problem
to be addressed is that $G_{\tau}$ necessarily has the
same rank  as $G$  as it contains $H'$ by virtue of by
\eIIIix\
and the fact that $T$ lies in $H'$.
Thus $G_{\tau}$ would therefore not relate simply to the
folded Dynkin diagram which has fewer
points than  this as also does the rank of ${\bf X}_n$ in
tables 1 and 2.

The resolution of the apparent paradox will be that
$G_{\tau}$ defined by \eIIIix\ is not semisimple. Rather
it is composed of $(k-1)$ invariant $ U(1)$ factors
times a semisimple factor which relates straightforwardly
to the folding when it is simple.

To understand this we need to establish the structure of
$T$, \etau\ in more detail. As $T \in H'$
\eqn\eIIIxi{T = e^{-2 \pi i Y.h}}
where $Y$ is a vector to be determined  and $(h_1, h_2,
\ldots, h_r)$  denote an orthonormal basis of the Cartan
subalgebra in apposition ${\bf h'}$ namely that
containing $E_1$,
\ete\ and generates $H'$. This basis is chosen to be
conjugate to
the basis of the original Cartan subalgebra
${\bf h}$, $(H_1, H_2,
\ldots, H_r)$,
\eqn\eIIIxii{h_i = P H_i P^{-1}, \hskip 1cm i=1, 2, \ldots,
r, \hskip .3cm  P \in G .}
Using this we can write $z(\tau)$ in \etsts\ also as
\eqn\eIIIxiia{z(\tau) = e^{-2 \pi i \l_{\tau(0)}^v . h},}
remembering that $z(\tau)$ lies in the centre of $G$ and
therefore commutes with $P$.
Since
\eqn\eIIIxiii{SY.hS^{-1} = \sigma(Y).h}
where $\sigma$ is the Coxeter element of $W(\bg)$ in the
form $\sigma_-\sigma_+$, \FLO , we can evaluate
$z(\tau)$,
\eIIIxiia ,  in another way, using \etsts\ and
\eIIIxi , and find

\eqn\eIIIxiv{z(\tau) = e^{-2\pi i(1 - \sigma)Y.h} .}
Comparing this two expressions, \eIIIxiia\ and \eIIIxiv, we
have
\eqn\eIIIxv{( 1-\sigma ) Y = \l_{\tau(0)}^v + \L_R(\bg^v)}
where the element of the coroot lattice $\L_R(\bg^v) $ is
undetermined.

As $(1 - \sigma)$ never vanishes, it has a unique inverse.
Further it maps the coweight lattice of $\bg$ into its
coroot lattice\FO by virtue of the identity:
\eqn\eIIIxvi{ \gamma_i = (\sigma - 1) \sigma^{-(1+c(i))/2}
\lambda_i}
Therefore,
\eqn\eIIIxvii{ T = e^{-2\pi i (1 - \sigma)^{-1}
\l_{\tau(0)}^v.h }  }
up to an element of $Z(G)$ dependent on the undetermined
element of the coweight lattice of $\bg$. This undetermined
factor is innocuous as it has no effect on the adjoint
action of $T$, \etsts.

Nevertheless we shall take advantage of this ambiguity in
the following way: replace $\l_{\tau(0)}^v $ in \eIIIxv\ by
$w(\l_{\tau(0)}^v )$ where $w \in W(\bg)$ is a Weyl group
element. Then, instead of \eIIIxvii,
\eqn\eIIIxviii{ T = e^{-2\pi i (1 - \sigma)^{-1}
 w \l_{\tau(0)}^v .h} }
where, by our previous comment, the dependence on $w$ is
carried by an innocuous central factor. The point is that
we shall find a choice of $w$, dependent on $\tau$, which
simplifies \eIIIxviii\ considerably.

\newsec{Concrete expression for $T$ and the structure of
$G_{\tau}$}

We now add the assumptions that $\bg $ is simply laced and
that
$\tau \in W_0(\bg)$ is direct. The latter condition excludes
the element of order 4 in $W_0$ $({\bf D}_{2n+1}) \cong Z_4 $.
The former condition means that all roots can be taken to have
length
$\sqrt{2}$. Then  fundamental weight $\l_i$ and fundamental
coweight $2\l_i/\a_i^2 = \l_i^v$ are identical. With these
assumptions we shall verify that there exists a choice of $w
\in W(\bg)$ such that
$T$ is given by \eIIIxi\   with
\eqn\eIVi{Y = (1-\sigma)^{-1} w\l_{\tau(0)} \equiv  wY'
            = {1 \over k} w \left( \l_{\tau(0)}
               + \l_{\tau^2(0)}
               + \cdots +  \l_{\tau^{(k-1)}(0)}\right)}
with this we can determine the structure of $G_{\tau}$,
\eIIIix , and relate it to the Dynkin diagram obtained by
folding
$\tri{\bg^{(1)}}$ with $\tau$.

Let us define
\eqn\eIVii{\sigma' = w^{-1} \sigma w}
which, being conjugate to $\sigma$, is equally a Coxeter
element of $W(\bg)$, but not the one whose action is induced
by $S$. Then \eIVi\ is equivalent to
\eqn\eIViii{Y' = (1-\sigma')^{-1}\l_{\tau(0)}
               = {1 \over k} \left( \l_{\tau(0)}
               + \l_{\tau^2(0)}
               + \cdots +  \l_{\tau^{(k-1)}(0)}\right) .}
This will be proven by constructing $\sigma'$ such that
\eqn\eIViv{\lt{p} = \left(1 + \s ' +{\s'}^2 + \cdots +
{\s'}^{p-1}\right)\lt{ }\; ,\hskip 2cm  p = 1, \ldots , k . }
Formally $\lt{k} = \l_0 = 0$ and this is guaranteed by
\eIViv\ if
\eqn\eIVv{(1-\s'^k) \lt{} = 0 .}
Unfortunately the verification of \eIViv\ and \eIVv\ has to
be  done on a case by case basis and is therefore relegated
to the appendix A. The case of $\bg = {\bf SU(mn)}$ and $\tau$
given by
\et\ is particularly instructive and easy to verify. It
follows from \eIViv\ that
\eqn\eIVvi{\s' \lt{p} = \lt{p+1} - \lt{}}
Summing from values of $p$ running from 1 to $(k-1)$ yields
$$
\s' Y' = Y' - \lt{ }
$$
which confirms \eIViii . Moreover, if we define
\eqn\eIVvii{ q'\left({ph(\bg) \over k}\right) = {1 \over k}
\sum_{m=1}^{k-1} e^{{-2\pi i pm \over k}} \lt{m}}
we likewise find
\eqn\eIVviii{\s' q'\left({ph(\bg) \over k}\right) =
 e^{{2\pi i p \over k}} q'\left({ph(\bg) \over k}\right) . }
As \eIVvii\ manifestly does not vanish, being composed of
$k$ linearly independent quantities, and as ${p \over k
}=\left({ph(\bg) \over k} \right){1\over h(\bg)}$ we deduce
that the $(k-1)$ integers ${ h\over k} ,{ 2h\over k} , \ldots
,  {(k-1)h\over k} $ are all exponents of $\bg$. Combined with
the statement that all the integers between 1 and $(h(\bg) -
1)$  coprime to $h(\bg)$ are also exponents, we have a
economical means of calculating the exponents of ${\bf E}_6,
{\bf E}_7$ and ${\bf E}_8$. Note that
\eqn\eIVix{ Y' + \sum_{p=1}^{k-1} q'\left({ph(\bg) \over
k}\right) = 0 .}
The structure of $G_{\tau}$, \eIIIx , can be determined by
considering the isomorphic group conjugate to $G_{\tau}$
within $G$
$$
G_{\tau}' = W^{-1} G_{\tau} W
$$
whose elements commute with
$$
T' = W^{-1} T W = e^{-2\pi i Y'.h}
$$
where $Y'$ was given in \eIViii . The generators of
${\bf g_{\tau}'}$ comprise the complete Cartan subalgebra in
apposition of
$\bg$, and step operators $F^{\a}$ for roots
$\a$ with respect to this Cartan subalgebra:
$$
F^{\a'} = W^{-1}F^\a W \sim F^{w^{-1}(\a)}
$$
which satisfy
\eqn\eIVx{ e^{-2\pi i Y' . w^{-1}\a } = 1 .}
But by \eIViii\ and the fact that the weights $\lt{p}$ are
all minimal so that for any root $\b$, $\lt{p} . \b = 0$ or
$ \pm 1$, we have
\eqn\eIVxi{ |Y'. \b | \leq 1 -{1\over k} < 1 .}
Hence the only solutions to condition \eIVx\  satisfy
\eqn\eIVxii{Y'\ .\  w^{-1}\a \; = \; Y\ .\ \a \;
= \; 0 .}
If $w^{-1} \a $ is positive this implies that it is
orthogonal to each of the minimal fundamental weights
$\lt{}, \lt{2}, \ldots , \lt{k-1}$.
This, in turn,  means that
\eqn\eIVxiii{ G_{\tau} = G_{\tau}^0 \otimes U(1)^{k-1} .}
where $G_{\tau}^0$ is semisimple. As $G_{\tau}$  has the same
rank as
$G$
\eqn\elVxiv{\hbox{rank}\, (G_{\tau}^0)=\hbox{rank}\, (G) + 1
- k.}
Furthermore, the
semisimple factor $G_{\tau}^0$ has a Dynkin diagram,
$\tri{{\bf g_{\tau}^0}}$,
 obtained by deleting from $\tri{\bg^{(1)}}$ the $k$
vertices $0, \tau(0), \ldots , \tau^{(k-1)}(0)$.
Thus, referring to Table 2, we see that deleting vertices
$0,2l-1$ from $\tri{{\bf D}_{2l}^{(1)}}$ yields $\tri{{\bf
A}_{2l-1}}$, deleting vertices $0,1$ from $\tri{{\bf
D}_{l+2}^{(1)}}$ yields $\tri{{\bf D}_{l+1}}$, deleting
vertices
$0,6$ from $\tri{{\bf E}_7^{(1)}}$ yields $\tri{{\bf E}_6}$,
while deleting vertices
$0,1,5$ from $\tri{{\bf E}_6^{(1)}}$ yields $\tri{{\bf D}_4}$.
Thus, in these cases, the Lie algebra $\bg_{\tau}^0$ is indeed
isomorphic to
${\bf X}_n$ defined by the first column in Table 2.
 In particular, we see that ${\bf g}_{\tau}^0$
is actually simple  in all cases except the final one
cited in Table 3 when deleting vertices $0,n,2n,\dots
(m-1)n$ from $\tri{{\bf A}_{mn-1}^{(1)}}$ yields $m$ copies of
$\tri{{\bf A}_{n-1}}$.

Notice also that it follows from \eIVxii\ that if $F^{\a}$
is a generator of ${\bf g_{\tau}^0}$
\eqn\eIVxiv{ w^{-1}\a . q'\left({ph\over k}\right) =
            \a . q\left({ph\over k}\right) = 0 , \hskip2cm
             p=1,2,...,k .}
Thus the roots of ${\bf g_{\tau}^0}$ are all perpendicular to
the
$(k-1)$ dimensional subspace spanned by the $(k-1)$
eigenvectors of
$\s$, (the Coxeter element induced by S) corresponding to the
$(k - 1)$ exponents $h(\bg)/k, 2h(\bg)/k\dots$. This was first
observed  by Dorey\Dorc\ for the particular case of
$\bg = {\bf E}_6$.

We shall now consider the action of $S$ on $G_{\tau}^0$.

\newsec{The twisted Coxeter element from the action of $S$ on
$G_{\tau}^0$}

We shall now assemble the preceding results. In section 3 we
saw that the conjugation by $S$ acted on $G_{\tau}$ as an
outer  automorphism. In section 4 we saw how $G_{\tau}$
factored into a semisimple group $G_{\tau}^0$ times an
Abelian group of dimension $(k-1)$ with generators
\eqn\eVoi{q\left( {h(\bg) \over k} \right)\!.h
\;\;,\;\; q\left( {2
h(\bg)
\over k}
\right)\!.h \;\; , \ldots , \;\; q\left( {(k-1) h(\bg) \over
k}
\right)\!.h .}
Because $ q\left( {p h(\bg) \over k} \right) $ are
eigenvectors of $\s $, $S$ acts on the Abelian factor
diagonally:
$$
Sq\left( {p h(\bg) \over k} \right).h S^{-1} =
\s q\left( {p h(\bg) \over k} \right).h =
e^{{2 \pi i p \over k}  }
 q\left( {p h(\bg) \over k} \right).h.
$$
As these eigenvectors are orthogonal to all the roots of
${\bf g_{\tau}^0}$, (4.15),
$S$ acts directly on ${\bf g_{\tau}^0}$. So
\eqn\eVi{S F^{\a} S^{-1} = F^{\s (\a)}. }
In fact, the roots of ${\bf g_{\tau}^0}$ fall into complete
orbits of
$h(\bg)$ elements under $\s$. This is because if one element
of an orbit is orthogonal to the eigenvectors of $\s$
mentioned above then so are all the other roots in the
orbit. If there are
$l$ such orbits then ${\bf g_{\tau}^0}$ has $lh(\bg)$ roots.
Thus
\eqn\eVii{lh(\bg) = r({\bf g_{\tau}^0}) h({\bf g_{\tau}^0})
= (r(\bg) + 1 - k) h({\bf g_{\tau}^0})}
equals the number of roots of ${\bf g_{\tau}^0}$ counted two
different ways. This makes it clear that the action of $\s$
on the roots of
${\bf g_{\tau}^0}$ differs from the action of its own
Coxeter  element. This different action is that of the twisted
Coxeter
element . By the
result of section 3,
$\s^k$ is inner as far as $G_{\tau}^0$ is concerned even
though $\sigma$ is
outer. Thus
$\s$, being an automorphism of the ${\bf g_{\tau}^0}$ root
system, is composed of a
$\tri{{\bf g_{\tau}^0}}$ diagram automorphism of order $k$
times a element of $W({\bf g_{\tau}^0})$. It therefore has
the same properties as the twisted Coxeter element
 of ${\bf X}_n\equiv {\bf g}_{\tau}^0$ described in chapters
8 and 14 of Kac' book \Kac and in \Spr
and can therefore be identified with it. In particular, the
twisted Coxeter element of ${\bf X}_n$
has the same order as the Coxeter element of ${\bf g}$,
namely $h({\bf g})$.

It can also be checked from \eVii\ that, when ${\bf
g_{\tau}^0}$ is simple, the Dynkin diagram
$\tri{{\bf X}^{(k)}_n} =
\tri{({\bf g_{\tau}^0})^{(k)}_n}$ has precisely $(l+1)$
vertices so that our two usages of the symbol $l$ indeed
agree.

By extension of the usual definition of an exponent, one
can define the twisted exponents of ${\bf g_{\tau}^0}$ as
being those powers of $\exp 2\pi i /h(\bg)$ which occur as
eigenvalues of
$\s$ applied to the Cartan subalgebra in apposition of
${\bf g_{\tau}^0}$. By our construction, the twisted
exponents are given by the set of exponents of ${\bf g}$ less
the
$(k - 1)$ exponents $h(\bg)/k , 2h(\bg)/k ,
\ldots , (k-1)h(\bg)/k$ associated with the abelian subalgebra
of ${\bf g_{\tau}}$, \eVoi. Again this
agrees with the results described in \Kac .

The action of $S$ applied to the Lie algebra  ${\bf
g_{\tau}^0}$ furnishes a grading of order $h(\bg)$. It is
precisely the grading \egra\ defined for ${\bf g}$ applied to
the subalgebra
${\bf g_{\tau}^0}$. When
${\bf g_{\tau}^0}$ is simple it coincides with the grading
defined in book of Kac by a different method and when ${\bf
g_{\tau}^0} $ is not simple (see Table 3) it provides an
interesting new
possibility.

It is this grading, the twisted principal grading, which
can be used to define the twisted affine Kac-Moody algebra
${\bf X}_n^{(k)}$. In the present manner of construction, the
natural basis for the algebra is a subset of the basis of
$\bg^{(1)}$  written in terms of its principal Heisenberg
subalgebra $\EM$ ($M$ equals an exponent of $\bg^{(1)}$) and
the quantities $\hat{F}(\a, z)$
ad-diagonalising the principal Heisenberg subalgebra. This
subset is
\eqn\eViii{
    \eqalign{&\EM : \hskip.7cm \hbox{ $M =$ a twisted
              exponent of ${\bf g_{\tau}^0}$ (mod $h(\bg)$),}
\cr
             &\hat{F}(\a , z) : \hskip.3cm \hbox{for
               roots $\a$ of $\bg$ satisfying $\a.
\qp{p} = 0; \; \; \; p = 1, 2, \ldots , (k-1)$ .}\cr}}
Since $\bg^{(1)}$ was simply laced, the roots $\a$ all have
the same length ($\sqrt{2}$ say) and this applies equally to
${\bf X}_n^{(k)}$. Thus, despite the Dynkin diagram (table 1)
not being simply laced, the algebra is nevertheless simply
laced in the sense just described. The most striking property
of ${\bf X}_n^{(k)}$ that  follows as a consequence is that
the vertex operator construction for
$\bg^{(1)}$ still applies to
${\bf X}_n^{(k)}$. This was originally demonstrated by
appealing to character formulae for the irreducible
representations \Kac\ but we shall see how it follows
naturally in our approach.

First we must consider how the fundamental highest weight
representations of ${\bf X}_n^{(k)}$ follows from those of
$\bg^{(1)}$ via folding. What happens is that the
inequivalent  fundamental representations of $\bg^{(1)}$ with
highest weights
$\L_{j},
\L_{\tau(j)}, \L_{\tau^2(j)}, \ldots, $ become identified as
a single fundamental representation of ${\bf X}_n^{(k)}$
whose highest weight is denoted $\L_{<j>} $. This is because
$X_{<i>} = \sum_{i \in <i>} X_i$ , $X_i = e_i ,f_i ,
h_i$ have the same action on each of these states and
they generate
${\bf X}_n^{(k)}$. Actually we expect the rest of the
principal Heisenberg subalgebra of $\bg^{(1)}$, namely $\EM$
for $M =
\left(n + {p \over k} \right) h(\bg), \;  n \in Z,
\;  p = 1,2, \ldots , (k-1)$ to be represented. As $m_j =
m_{\tau (j)}$ the level of $\L_{<j>}$ is the same as the
levels of $\L_j, \L_{\tau(j)}, \L_{\tau^2(j)}, \ldots \; .$
Recall that as $ \bg^{(1)}$ is simple laced
\eqn\eViv{\hat{F}(\a,z) ^{m_j +1} = 0}
in the $\bg^{(1)}$ irrep with highest weight $\L_j$. This
therefore remains true in the $\L_{<j>}$ irrep of
${\bf X}^{(k)}_n$ for the surviving $\hat{F}(\a,z)$, namely
those satisfying \eViii . Furthermore,
$\hat{F}(\a,z)^{m_j} / m_j !$ is a vertex operator\KOa , that
is a normal ordered exponential of the principal Heisenberg
subalgera $\EM $

Without loss of generality we can replace $\a$ by
$\gamma_i$ the standard representative on its orbit under
the action of $\s$ as explained in \OTUa . Then if
$\hat{F}^i(z) \equiv \hat{F}(\gamma_i , z)$ in the
representation with highest weight $\Lambda_j$, we
have the generalised vertex operator construction\KOa ,
\eqn\eVv{{ \hat{F}^i(z)^{m_j} \over m_j ! } =
e^{- 2 \pi i \l_i . \l_j} Y^i Z^i ,}
where
\eqn\eVvi{ Y^i = \exp \sum_{M>0} {\gamma_i . q([M]) z^M
\hat{E}_{-M} \over M} , \hskip 1cm
Z^i = \exp \sum_{M>0} -{\gamma_i . q([M])^* z^{-M}
\hat{E}_{M} \over M} .}
In order to establish that \eVv\ and \eVvi\ makes sense in
${\bf X}_n^{(k)}$ as well as $\bg^{(1)}$ we need to check two
things.

First consider the sum in the exponential. In $\bg^{(1)}$ the
sum includes all the exponents of $\bg^{(1)}$ but in
${\bf X}_n^{(k)}$  the sum should be restricted to the
twisted exponents of ${\bf X}_n^{(k)}$. This restriction is
automatically guaranteed as when $\fiz$ lies in ${\bf
X}_n^{(k)}$,
$\g_i$ satisfies condition \eViii . Thus the surviving
$\fiz \in {\bf X}_n^{(k)}$ are obtained by exponentiating
the twisted principal Heisenberg subalgebra ones.

The second point concerns the phase factor in \eVv \KOa\
which we know plays an important r\^ole in determining the
asymptotic behaviour of the soliton solution. We have to
show that this phase is unaltered if $\l_j$ is replaced by
$\l_{\tau(j)}$ so that it is independent of the
representative $\L_j$ or $\L_{\tau(j)}$ chosen for
$\L_{<j>}$. Consider the expectation value
\eqn\eVvii{\lltj \;
{\left(\hat{F}^i(z)\right)^{m_j} \over m_j ! } \;
\rltj = e^{-2\pi i \l_i . \l_{\tau(j)}} .}
Lifting $\tau$ to the outer automorphism $\hat{\tau}$ of
$\bg^{(1)}$ expression \eVvii\ equals
\eqn\eVviii{\llj \;
{\left(\hat{\tau}^{-1}\left(\hat{F}^i(z) \right)
\right)^{m_j}
\over m_j !}\;\rlj .}
But $\hat{\tau}^{-1}\left(\hat{F}^i(z)\right) =
\hat{F}^i(z)$ precisely when
$\hat{F}^i(z) \in {\bf X}^{(k)}_n$. Thus this expression
equals
expression \eVvii\ with $\tau(j)$ replaced by $j$ and the
desired conclusion follows.

\newsec{ Particles and solitons in twisted affine Toda field
theories }

We can now use the results of the preceding work to determine
the mass spectrum (and other properties) of both the quantum
particles and the classical soliton solutions in the twisted
affine Toda field theories. The results are quite simple and
further indicate that the twisted theories more resemble the
simply-laced untwisted theories than the non simply laced
untwisted theories despite having non-simply laced Dynkin
diagrams.

In the simply-laced untwisted theories based on $\bg^{(1)}$
with which we start there is a one to one correspondence
between the $r \equiv $ rank $\bg$ species of quantum
particles and the same number of species of classical soliton
solution. Furthermore the ratio between the corresponding
masses is given by
\eqn\eVlo{{\hbox{Mass}(\hbox{Soliton } i)
\over\hbox{ Mass}(\hbox{Particle } i)}=
{4h(\bg)\over|\b^2|\hbar\gamma_i^2}.}
Therefore, for the simply laced theories, this number is
 universal in that it is independent of the
species $i$ in question.
 This was discovered for ${\bf A}_n$ by Hollowood, \Hola,
and later generalised to other untwisted
theories \OTUb, \MM , \ACFGZ .  When such a theory is folded
to give a twisted theory, subsets of the quantum particles
and classical solitons survive preserving this
correspondence. The same mass values survive unchanged,
thereby maintaining the universal ratio \eVlo, independent
of $i$, but with
$h(\bf g)$ equalling the twisted Coxeter number of
${\bf g}_{\tau}^0\equiv {\bf X}_n$.

On the other hand, when such a theory is folded to give an
untwisted nonsimply-laced theory the result is somewhat
different.  Subsets of quantum particles and classical
solitons again survive, and can still be put into
correspondence. However  some of the soliton masses will
change so that the mass ratio is still given by \eVlo but
no longer universal as the squared lengths of the roots
$\g_i$ differ. This is the result of \OTUb\ which  will be
confirmed below by a simpler argument.

To see all this in greater detail, we recall that the quantum
particle of species $i$ in the $\bg^{(1)}$ theory is
associated with the orbit under the action of $\sigma$  of
the $h(\bg)$ roots  $\gamma^i , \sigma \gamma^i,
\ldots $ through the corresponding step operators of the
finite dimensional Lie algebra $\bg$, $F^{\g^i}$ etc. A
similar association applies to the classical soliton species
except that, according to the explicit construction, the
correspondence is with the generators of the affine Kac-Moody
algebra, $\hat F^{\g^i}(z), \hat F^{\sigma (\g^i)}(z) =
\hat F^{\g^i}(ze^{{-2\pi i\over h}}) \ldots$. Under the
folding, the generators surviving to the subalgebra $\bgt^0
\in \bg$ or to its affinisation are, by definition, those
invariant under the lift of $\tau
\in W_0(\bg )$. These are determined by the result of \OTUb :
$$
\eqalign{\tilde{\tau}( F^{\g^i}) &= e^{-2\pi i \l_i .
      \l_{\tau(0)}}  F^{\g^i} \cr
\hat{\tau}(\hat{F}^{\g^i}(z)) &= e^{-2\pi i \l_i .
      \l_{\tau(0)}} \hat{F}^{\g^i}(z) \cr}
$$
and the fact that, if $F^{\g^i}$ survives, so do all the
step operators for the roots on the  $\sigma$ orbit of
$\gamma^i $. Thus the condition for the survival  under
folding of either a quantum particle of  species  $i$ or a
soliton of species  $i$ is exactly the same:
\eqn\eVIi{ e^{-2\pi i \l_i .\l_{\tau(0)}} = 1}
It is easy to evaluate this condition. Using the numbering
of table 2, the results are
$$\eqalignno{ {\bf E}_6^{(1)} \ (\tau(0) = 1) :& \ i = 3,6 \cr
{\bf E}_7^{(1)} \ (\tau(0) = 6 ) :& \ i = 1,2,3,5 \cr
{\bf D}_{l+2}^{(1)} \ (\tau(0) = 1 ) :& \ i = 1,2, \ldots ,l
\cr
{\bf D}_{2l}^{(1)} \ (\tau(0) = 2l ) :& \ i =2,4,\ldots,
2l-2,2l \quad  \hbox{for $l$ even}\cr
                                    & \  i =2,4,\ldots, 2l-2,
2l-1 \quad \hbox{for $l$ odd }\cr}
$$
and agree with the calculations of Braden et al \BCDSb\ for
the quantum particles for the twisted theories.

We saw in section 4 that the condition \eVIi\ for survival
could be written in another way, eq. \eIVxiv\ or \eViii . The
significance of this presentation, namely that the root to
the surviving generators be orthogonal to the space spanned
by $         q\left( h(\bg)/k \right) , q\left(2h(\bg)/k
\right) ,
\ldots$,   is that the subset of surviving particles (be they
quantum particles or solitons) is closed, and hence self
consistent, under the operations of

i. antiparticle conjugation,

ii. fusing.

\noindent (i) follows because the orbit associated with the
antiparticle of species $i$ consists of the negatives of the
roots in the orbit containing $\g_i$ and so if condition
\eIVxiv\ is satisfied by the particle it is satisfied by the
antiparticle. (In fact, the surviving species equal their
anti-species). If species $i$ and $j$ fuse
to give
$k$, then, by Dorey's rule, $\g_k$ can be expressed as the
sum of two
roots in the orbits containing $\g_i$ and
$\g_j$ respectively. Evidently if orbits $i$ and $j$ satisfy
\eIVxiv\ so does species $k$. Thus the spectrum (of either
quantum particles or solitons) of the  twisted affine Toda
field theories will contain  it own antiparticles and will
still couple by means of Dorey's fusing rule. This means
that, in the quantum theory, the S-matrix  in the untwisted
theory, restricted to the quantum particles states of the
twisted theory, will satisfy unitarity, crossing and
bootstrap property, as argued in the case of ${\bf
D_4}^{(3)}$ by Fring and Koberle \FKa . However, as Dorey
\Dora \Dorb\
has noted, there may be doubt about the positivity
properties. Certainly the quantum corrections to the same
quantum particle mass will differ in the two theories due to
the different spectrum of intermediate states allowed.

There has been much discussion of the proposal that the
quantum twisted theory be related to the untwisted non simply
laced theories whose Dynkin diagram is obtained by reversing
all the arrows \DGZ, \Dorc, \CDS. It is  tantalising that
these two theories do share isomorphic Poincar\'e algebras
\eIi\ as their generalised exponents coincide \Kac\
(Corollary 14.3).

Whatever the kind of folding involved, as long as it is
direct, it was shown in section 2, that the masses of the
surviving quantum particles are unchanged in the folding but
we still have to check  what happens to the classical soliton
masses, or indeed the multisoliton solutions more generally.
First we need to consider the expectation values $M_i$
and check how the condition
$M_i = M_{\tau(i)}$ of section 2, needed to define the
corresponding quantities
$M_{<i>} = M_i$ for the twisted theory, is satisfied whenever
$\tau$ is a direct symmetry of
$\tri{\bg^{(1)}}$. Recall that for a multisoliton solution of
the $\bg^{(1)}$ theory
\eqn\eVIii{M_j = \langle \Lambda_j | g(t) |\Lambda_j
\rangle }
where
$$\eqalignno{g(t)& = V(t) \  g(0) \  V(t)^{-1} ,\cr
             g(0)& = \prod_j \exp \left[ Q_{n(j)}
\hat{F}^{n(j)}(z_{(n(j)})  \right], \cr
             V(t)& =
\exp\left(\mu \hat{E}_{-1}t_{-1}\right)
\exp\left(-\mu  \hat{E}_1t_1\right),\cr }
$$
{}From this we see that the symmetry condition
$M_i=M_{\tau(i)}$ is satisfied whenever the group element
satisfies
$\hat\tau(g(0))=g(0)$ i.e. it is generated by the
folded algebra. If $\tau\in W_0(\bg)$, this means that
the species of $\hat{F}$ satisfy \eVIi , or, equivalently,
(5.4).

If, instead, $\tau \in $ Aut$\tri{\bg}$, so that the folding
leads to a non simply laced untwisted theory, the
condition is again satisfied but in a somewhat different way.
In this case it was shown in \OTUb\ that the surviving
$\hat{F}^{<i>}(z)$ were linear combination of the original
ones:
\eqn\eVIiii{\hat{F}^{<i>}(z) = \sum_{i\in <i>}\hat{F}^{i}(z) }

We shall see that this implies that the single soliton of
species $<i>$ of the folded theory will arise from
folding a special $|<i>|$-soliton solution of the unfolded
theory.

 We have been using the result of section 2 whereby,  given a
``direct" symmetry of the Dynkin diagram $\tri{\bg^{(1)}}$,
the symmetric solution of the unfolded theory yields solutions
of the folded theory and vice-versa. In order to understand
the resultant soliton masses we shall present a similar
theorem for the Lagrangian and for the energy momentum
tensor: that a symmetric configuration for the unfolded
theory (not necessarily a solution) yields a configuration of
the folded theory such that the respective Lagrangian (and
energy momentum tensor) assume identical values.

In order to see this we first note that the folding of the
Cartan matrix \eK\ is satisfied by folding the roots as
follows:
\eqn\eVIiv{\a_{<i>} = \sum_{i\in <i>} {\a_i \over |<i>|} ,}
where $|<i>|$ is the number of roots in the orbit of $i$. It
follows that
\eqn\eVv{ {2m_i\over \a_i^2} = {2m_{<i>}\over |<i>| \
\a_{<i>}^2} }
so that the constraint condition \eefi\ is preserved. We now
apply this to the interaction  term in the Lagrangian (and the
energy momentum tensor)
\eqn\eVIvi{V(\phi) = {2 \mu^2 \over \b^2} \sum_{i=0}^r {m_i
\over
\a_i^2 }\left( e^{\b \phi_i} -1 \right) }
and see immediately that for symmetric  configurations it
equals
\eqn\eVIvii{{2 \mu^2 \over \b^2} \sum_{<i>}^r {m_{<i>} \over
\a_{<i>}^2 }\left( e^{\b \phi_{<i>}} -1 \right) }
which is the corresponding interaction term for the folded
theory. The same result applies to the kinetic terms
$\partial_{\mu} \phi . \partial^{\mu} \phi $ and
$\partial_{\mu} \phi . \partial_{\nu} \phi $ in the
Lagrangian and energy-momentum tensor but the proof is more
complicated and relegated to  appendix 2. The main difficulty
is to express the kinetic term in term of the variables
$\phi_i$, \ephi , in a way which manifestly respects the
symmetry under Aut$\tri{\bg^{(1)}}$.

In particular, if we have a symmetric multisoliton solution
of the unfolded theory, it will yield a multisoliton solution
of the folded theory with the same energy and momentum.
Nevertheless we cannot conclude that the solitons have the
same masses in the folded and unfolded theory as soliton
number is not conserved necessarily preserved in the folding
as we now see.

In the case that $\tau \in W_0(\bg)$, so that the folded
theory is twisted, a multisoliton solution of species $n(1) ,
n(2) , \ldots , n(k)$  satisfying \eVIi\  is symmetric and
survives as a soliton with the same interpretation. According
to the result of \OTUa,  its energy and momentum is given by
$$\sqrt{2} P^{\pm} = \sum_{i=1}^k M_{n(i)} e^{\pm
\eta_{n(i)}}
$$
As this result applies to the folded theory also, by our
theorem, the folded soliton solution therefore possesses the
same mass as the unfolded one. This is the result mentioned at
the beginning.

On the other hand, if $\tau \in $ Aut $\tri{\bg}$ so that the
folded theory is untwisted but non simply laced, the
symmetric solution could involve an exponential of \eVIiii
\eqn\eVIvii{e^{Q \hat{F}^{<i>}(z) } = \prod_{i\in <i>} e^{Q
\hat{F}^{i}(z)} .}
The left hand side would create a single soliton of the
folded theory but the right hand side would create a
superposition of $|<i>|$ solitons of the unfolded theory all
with the same coordinate and rapidity. We have used the fact
that the pieces in \eIViii\ mutually commute since they have
same rapidity.

When we equate the consequent contributions to the energy and
momentum tensor of the soliton we have
$$
\sqrt{2} P^{\pm} = M_{<i>} e^{\pm \eta} = \sum_{i \in <i>}
M_{i} e^{\pm\eta} .
$$
Hence we conclude that the mass of the soliton of species
$|<i>|$,
$$ M_{<i>} =  |<i>| M_{i}
$$
(as $M_i = M_{\tau(i)} $). This confirms the result \eVlo\ of
\OTUb\  with the bonus of a more physical understanding. Note
that this explains why MacKay and McGhee \MM\ overlooked some
of the soliton  species in the untwisted nonsimply-laced
theories. They  considered folding only single soliton
solutions and so ignored configurations such as \eVIvii .

\bigbreak\bigskip\bigskip\centerline{{\bf
Acknowledgements}}\nobreak We are grateful for discussions
with H. Braden, E. Corrigan, P. Dorey and J.W.R. Underwood.
Marco Kneipp wishes to thank CNPq (Brazil) for financial
support.

\bigbreak
\bigskip
\bigskip

\ni{\bf Appendix A:  Element $T$ of $G$ corresponding to
$\tau \in W_0({\bf g})$ }
\medskip

The lift of the diagram automorphism $\tau \in W_0(g)$ is
provided by the inner automorphism \etau\ in which $\tau$ has
the form $\exp (-2\pi i Y \cdot h )$, \eIIIxi . We shall now
verify, on a case by case basis, the existence of an element
$w$, of the Weyl group, $W(g)$, such that $Y$ has the form
\eIVi . It is instructive to start with the case ${\bf g} =
{\bf A}_{mn-1}$ in Table 3. As this is the only case in
which $\tau$ may have an order greater than 3, namely $m$,
it is the most complicated.

We recall that the $mn-1$ fundamental weights $\l_l$ can be
expressed in terms of $mn$ unit vectors $\epsilon_i$:
$$
\l_l = \sum_{i=1}^l \e_i - {l\over mn} \sum_{i=1}^{mn} \e_i
\ ,
\ \ \ \ \ l = 1,2, \dots , mn-1 \ .
$$
The Weyl group, $W({\bf A}_{mn-1})$, is isomorphic to the
group of permutation of the unit vectors, any element can
be denoted by the standard permutation notation. The
following element cyclically permutes the $mn$ unit vectors
and hence is a Coxeter element
$$
\eqalignno{\sigma' = & \left( 1\ ,n+1\ ,2n+1\ , \dots , (m-1)n
+1\ , 2\ , n+2\ , 2n+2\ , \dots ,(m-1)n + 2\ , \dots ,\right.
\cr & \left.  n\  , n+n\ , \dots ,
(m-1)n + n\  \right) \ . }
$$
It is easy to check that $\sigma'$ satisfies \eIViv\ and
\eIVv\ so that the desired result follows.

Now we turn to the four cases in Table 2. When $\tau$ has
order 2, as it does unless ${\bf g} = {\bf E}_6$, all that
has to be shown is the existence of a Coxeter element
$\sigma'$ conjugate to a standard one $\sigma$ satisfying
$$
\sigma' \l_{\tau(0)} = - \l_{\tau(0)} \ .
$$

Let us now consider ${\bf D}_r$ with its simple roots
constructed out of the unit vectors in the usual way
$$
\eqalignno{\alpha_i =& \e_i - \e_{i+1} \ , \  \ i = 1, \dots
, r-1 \cr
\alpha_r =& \e_{r-1} + \e_r }
 $$
so that the fundamental weights are
$$
\eqalignno{\l_i = & \sum_{i=1}^r \e_i \ \ \ \ \ \ \ \ i= 1,
2,
\dots ,r - 2 \cr
\l_{r-1} = &{1 \over 2} \left( \sum_{i=1}^{r-1} \e_i - \e_r
\right) \  , \ \ \ \ \ \l_r = {1 \over 2} \sum_{i=1}^r \e_i
.}
$$
Then it is easy to see that the Coxeter element $\s'' = \s_1
\s_2 \cdots \s_r$ (with $\s_i$ the reflection in $\a_i$)
has the following action on the unit vectors
$$
\eqalignno{\e_1 &\rightarrow \e_2 \rightarrow \dots
\rightarrow \e_{r-1} \rightarrow -\e_1 \rightarrow -\e_2
\rightarrow \dots \rightarrow -\e_{r-1} \rightarrow \e_1 \ ,
\cr
\e_r& \rightarrow -\e_r \ . }
$$
Hence as $\s_{\e_1 + \e_r} (\e_r) = -\e_1$
$$
\s' = \s_{\e_1 + \e_r} \ \s'' \ \s_{\e_1 +\e_r}
$$
has the action $\s' \l_1 = -\l_1$. This is the desired
result if $\tau(0) = 1$ as in the second line of Table 2.

If now $r$ is even, equal to $2l$, say, $\sigma''$ also
reverses the signs of $\e_1 - \e_2 + \e_3 - \e_4 + \cdots +
\e_{2l-1} $. Hence it reverses the sign of one half this
plus or minus $\e_{2l}$.   But this is a spinor weight Weyl
conjugate to $\l_{2l}$ or $\l_{2l-1}$ according to the sign
chosen. This establishes the existence of $\s'$ reversing
the sign of $\l_{2l-1}$ as needed when $\tau(0) = 2l-1$ as
in the first line of Table 2.

For ${\bf E}_6$ let us consider the element of the Weyl
group
$$
w = \s_\beta \s_{\a_1} \s_{\a_5} \ , \ \ \ \  \
\beta \equiv \a_1 + 2\a_2 + 2\a_3 + \a_4 + \a_6 \ .
$$
One can check that
$$
\eqalignno{w \l_1 =& -\l_1 + \l_5 \ , \cr
           w \l_5 =& -\l_2 + \l_4 \ .}
$$
Now using the bicolouration such that $c_1 = c_5 = -1$ and
the identities \FO
$$
\eqalignno{\s \l_{i_-} =& \l_{i_-} - \a_{i_-} \ , \cr
           \s_\pm \a_{i_\mp} =& \a_{i_\mp} - K_{i_\mp j}
\a_j \ , \cr
           \s_\pm \a_{i_\pm}  =& - \a_{i_\pm} \ ,  }
$$
it can be proven that
$$
\eqalignno{\left(1 + \s \right)  w\l_1 =& w\l_5 \ , \cr
           \left(1 + \s +\s^2 \right) w\l_1 =& 0 \ ,}
$$
which correspond to \eIViv\ for $\tau(0) = 1$.

Finally, for ${\bf E}_7$, let $w = \s_\beta$ , where
$\beta \equiv \a_1 + 2\a_2 + 3\a_3 + 2\a_5 + \a_6 + \a_7$.
Then,
$$
w \l_6 = -\l_4 + \l_6 + \l_7 \ .
$$
Now, like before, using  the bicolouration such that $c_4 =
c_6 = c_7 = -1$ it is easy to prove that
$$
(1 + \s) w\l_6 = 0 \ .
$$
This completes the claimed result.

\bigbreak
\bigskip
\bigskip

\def\ni{\noindent}
\def\t{\tau}
\def\m{\mu}
\def\r{\rho}
\def\tri#1{\bigtriangleup(#1)}
\def\a{\alpha}
\def\l{\lambda}
\def\bg{\bf g}

\ni{\bf Appendix B: Folding of the energy momentum tensor}
\medskip

Here we complete the argument of section 6 and show
that, for a field configuration symmetric under $\t\in
Aut\tri{\bg^{(1)}}$, the energy momentum tensors of the folded
and unfolded affine Toda field theories are equal, providing
$\t$ is direct.

Given the Lie algebra $\bg$, we define the quantities
$$
\m_i=\l_i^v-{2m_i\,\r^v\over\a_i^2\, H}\qquad i=0,1,2\dots
r\eqno(B1)
$$
where $\l_1^v,\l_2^v,\dots \l_r^v$ are the
fundamental coweights, $\r^v$ their sum, while
$\l_0^v$ vanishes. The number
$$
H\equiv\sum_{i=0}^r{2m_i\over\a_i^2}.\eqno(B2)
$$
If the long roots have length $\sqrt2$, $H$ denotes either the
Coxeter number or twisted Coxeter number, whichever is
relevant. Then, because of the constraint
\eefi , the field
$\phi$ can be written
$$
\phi=\sum_{i=0}^r\phi_i\m_i.\eqno(B3)
$$
The significance of
this is that it provides the way of introducing all $r+1$
variables $\phi_0,\phi_1,\dots\phi_r$, \ephi , which will
respect the full symmetry of the extended Dynkin diagram,
$Aut\tri{\bg^{(1)}}$. To see this, first note that it is easy
to check that
$$
\a_i\cdot\m_j=\delta_{ij}-{2m_i\over H\a_i^2}\qquad
i,j=0,1,\dots r.\eqno(B4)
$$
Since
$$
\t(\a_i)\equiv\a_{\t(i)}$$ defines the linear map $\t$,
given
$\t\in Aut\tri{\bg^{(1)}}$, we can easily check that
$$
\a_i\cdot\t(\m_j)=\a_i\cdot\m_{\t(j)}, \eqno(B5)
$$
and hence
$$
\m_i\cdot\m_j=\m_{\t(i)}\cdot\m_{\t(j)}.\eqno(B6)
$$
So the term $\partial_{\m}\phi\cdot\partial_{\nu}\phi$ in the
energy momentum tensor can be written
$$
\sum_{i.j=0}^r\m_i\cdot\m_j\,\partial_{\m}\phi_i
\partial_{\nu}\phi_j,
\eqno(B7)
$$
a form which explicitly respects all the symmetries of the
extended Dynkin diagram.

Using \eVIiv\ one sees that $H$ is unchanged by the folding if
it is direct. Hence, by (B4),
$$
\m_{<j>}=\sum_{j\in<j>}\m_j.\eqno(B8)
$$
Hence, for
symmetric field configurations, expression (B7) equals
$$
\sum_{<i>,<j>}\m_{<i>} . \m_{<j>}\partial_{\m}\phi_{<i>}
\partial_{\nu}\phi_{<j>}.
$$
the same argument evidently applies to the kinetic term in
the Lagrangian,
${1\over2}\partial_{\m}\phi\cdot\partial^{\mu}\phi.$

\listrefs

\bye